\documentclass[manuscript]{emulateapj}
\usepackage[footnotes]{trackchanges}
\usepackage{natbib}
\usepackage{times}

\begin{document}

\title{A forecast for the detection of the power asymmetry from galaxy surveys}

\author{
Zhongxu~Zhai\altaffilmark{1},
Michael~R.~Blanton\altaffilmark{1}
}
\altaffiltext{1}{Center for Cosmology and Particle Physics, Department of Physics, New York University, 4 Washington Place, New York, NY 10003, USA.}

\begin{abstract}
We investigate the possibility of detecting in redshift surveys a hemispherical power asymmetry similar to that first reported in CMB observations. We assume the hemispherical asymmetry arises from a  linear gradient in comoving coordinates in the perturbation amplitude. We predict the resulting clustering of galaxy or galaxy cluster tracers using an excursion set approach; doing so accounts for the variation of both the underlying clustering and the tracer bias. Based on the predicted variation of the clustering of tracers, we perform a  Fisher matrix forecast of the galaxy clustering amplitude and  calculate the statistical significance for ideal surveys and  planned surveys. The results indicate that the DESI galaxy survey would be able to detect this signal with higher than $3\sigma$  significance if the asymmetry does exist. We also investigate the  amplitude  and scale dependence of the above result. The DESI  galaxy survey can probe the dipole amplitude higher than 0.04, which correspond to a $\pm4\%$ difference of the temperature fluctuation along and opposite the dipole direction, at least at the $2\sigma$ level. Additionally, we investigate a modulation of the power spectrum that exhibits asymmetry only for large scales. This modulation is potentially detectable. For Milky Way galaxy mass tracers, the scale-dependent modulation yields a larger change in the large scale power spectrum than does a scale-independent modulation, because the former does not alter the bias.
\end{abstract}


\section{Introduction}

The standard cosmological model is based on the concept of inflation, which hypothesizes that the universe went through a period of accelerated expansion at very early times (\citealt{Guth_1981, Linde_1982, Bassett_2006, Linde_2008}). This model explains the origin of the structure in the universe, and predicts that the energy density perturbations are distributed as a statistically homogeneous and isotropic Gaussian random field. Therefore it is anticipated that the temperature and polarization anisotropies of the Cosmic Microwave Background (CMB), which are related to the perturbations through linear theory, should be close to a statistically isotropic and homogeneous Gaussian random field in a similar way. 

However, after the CMB maps from Wilkinson Microwave Anisotropy Probe (WMAP) were released, many detailed studies reported evidence for unexpected features of the microwave sky, which if real would indicate the violation of statistical isotropy and Gaussianity. These anomalies include the smallness of the quadrupole moment (\citealt{deOliveira_2004}), alignments of low multipole moments (\citealt{deOliveira_2004, Copi_2004, Schwarz_2004, Land_2005}), special cold spots (\citealt{Vielva_2004, Cruz_2006, Cruz_2007}), and the hemispherical power asymmetry (\citealt{Eriksen_2004}). The latest investigations of these CMB anomalies are examined in great detail by \cite{Planck_2016}, and also reviewed by \cite{Schwarz_2016}.

In this paper, we focus on the hemispherical power anomaly which suggests a direction dependence of the CMB angular power spectrum, first reported by \cite{Eriksen_2004}, and further confirmed and revisited by many others (\citealt{Hansen_2004, Eriksen_2005, Eriksen_2007, Hansen_2009, Hoftuft_2009, Flender_2013, Aiola_2015}). This  power asymmetry can be modeled as a dipolar modulation $\frac{\Delta T}{T}(\hat{\mathbf{e}})=\frac{\Delta T}{T}_{0}(\hat{\mathbf{e}})[1+A\hat{\mathbf{e}}\cdot\hat{\mathbf{p}}]$, where $\frac{\Delta T}{T}(\hat{\mathbf{e}})$ and $\frac{\Delta T}{T}_{0}(\hat{\mathbf{e}})$ are the modulated and unmodulated temperature fluctuation respectively in direction $\hat{\mathbf{e}}$, and $A$ and $\hat{\mathbf{p}}$ are the dipolar modulation amplitude and direction (\citealt{Prunet_2005, Gordon_2005}). This power asymmetry feature has been examined with different methods and algorithms. The results are found to be qualitatively consistent with a detection significance of about $3\sigma$, including a direct comparison of the power spectral of the CMB map with different positions and radii of the angular disks (\citealt{Hansen_2004}), topological test of the CMB anisotropy by measuring the genus (\citealt{Park_2004}), application of the BiPolar Spherical Harmonics techniques (\citealt{Hajian_2003, Hajian_2006}), a direct likelihood fit of the dipole model (\citealt{Eriksen_2007}), usage of a local variance estimator (\citealt{Akrami_2014})  and so on.
The latest results from the Planck 2015 release using the \texttt{Commander} map finds that the greatest power is in the direction $\hat{\mathbf{p}}=(l, b)=(230^{\circ}, -16^{\circ})\pm24^{\circ}$ with an amplitude of $A=(0.066\pm0.021)$ (\citealt{Planck_2016}). 

Since the first reports of this signal, several theories have
been proposed to explain it. For instance, 
\cite{Erickcek_2008a, Erickcek_2008b} proposed a model involving 
a large amplitude superhorizon perturbation to the 
curvaton field (\citealt{Mollerach_1990, Linde_1997, Moroi_2001, Lyth_2002}), 
which has a statistically inhomogeneous amplitude. The 
effects of this superhorizon fluctuation may leave footprints as an anisotropic distribution of the large scale structure on the sky (\citealt{Schwarz_2016}), but the explorations with quasars and other tracers found that these were consistent with a statistical isotropic expectation (\citealt{Hirata_2009, Pullen_2010, Gibelyou_2012}).
In particular, using the number density of quasars, \cite{Hirata_2009} found an upper limit on the dipole of no more than 0.027 at the 99\% confidence level. This implies that if the Planck result is correct and not a statistical fluke, the dipole would have to be scale-dependent.

As pointed out by \cite{Hirata_2009}, any early-Universe explanation for the hemispherical asymmetry is likely to produce a spatial gradient in $\sigma_{8}$, the RMS fluctuation of matter in 8 $h^{-1}$Mpc spheres at $z=0$. One consequence of such a variation is that the number density and clustering of the galaxy or other tracers may be higher in some regions than others. Although the studies using number densities have not found statistically significant signals of a power asymmetry (\citealt{Hirata_2009}), we note that the spatial gradient of $\sigma_{8}$ also leads to different clustering amplitudes of the large scale structure in two different directions if the hemispherical power asymmetry found by CMB does exist. Therefore in this paper, we focus on the question whether it is possible to verify or falsify the existence of a hemispherical power asymmetry by measuring the clustering amplitude of the large scale structure from surveys of galaxies or galaxy clusters. We consider ideal surveys of dark matter halos, and then specifically consider the Dark Energy Spectroscopic Instrument (DESI,  \citealt{DESI_2016}) galaxy survey and the Extended ROentgen Survey with an Imaging Telescope Array (eROSITA, \citealt{eROSITA_2012}) cluster survey.

Our paper is organized as follows. Section \ref{sec:model} briefly describes the basics of the extended Press-Schechter formalism for predicting the halo mass function and halo bias. The methodology used to estimate the measurement uncertainty in the clustering amplitude is presented in Section \ref{sec:ideal}, along with the application to an ideal survey. In Section \ref{sec:real}, we apply our analysis to project the constraining power of actual surveys of DESI and eROSITA. Finally, we discuss our findings in Section \ref{sec:conclusions}. Throughout this paper, we assume a fiducial cosmological model from \cite{Planck_2016_pars}: ($\Omega_{m}$, $\Omega_{b}$, $h$, $n_{s}$, $\sigma_{8}$) = (0.308, 0.048, 0.678, 0.968, 0.829).

\section{Model for clustering and number density}
\label{sec:model}

The halo mass function (hereafter HMF) defines the comoving number density of dark matter haloes in the universe at a given redshift as a function of their mass. Its differential form can be written as
\begin{equation}\label{eq:hmf}
\frac{dn}{d\log{M}} = \frac{\rho_{0}}{M}f(\sigma) \Big| \frac{d\ln{\sigma}}{d\ln{M}}\Big|,
\end{equation}
where $\rho_{0}$ is the mean matter density of the universe, and $\sigma$ is the mass variance defined as
\begin{equation}
\sigma = \frac{1}{2\pi^{2}}\int P(k,z)\hat{W}^2(k,R)k^2dk,
\end{equation}
where $P(k,z)$ is the matter power spectrum at redshift $z$ and $\hat{W}$ is the Fourier transform of the top-hat window function at radius $R$. The form of $f(\sigma)$ defines a particular HMF. In our calculation, we consider the extended Press-Schechter (EPS) formalism (\citealt{PS_1974, EPS_1991}) which predicts
\begin{equation}
f(\sigma)=\sqrt{\frac{2}{\pi}}\frac{\delta_{c}}{\sigma}\exp{\Big[-\frac{\delta_{c}^2}{2\sigma^2}\Big]},
\end{equation}
where $\delta_{c}=1.686$ is the critical overdensity for collapse.

\begin{figure}[htbp]
\begin{center}
\includegraphics[width=9cm]{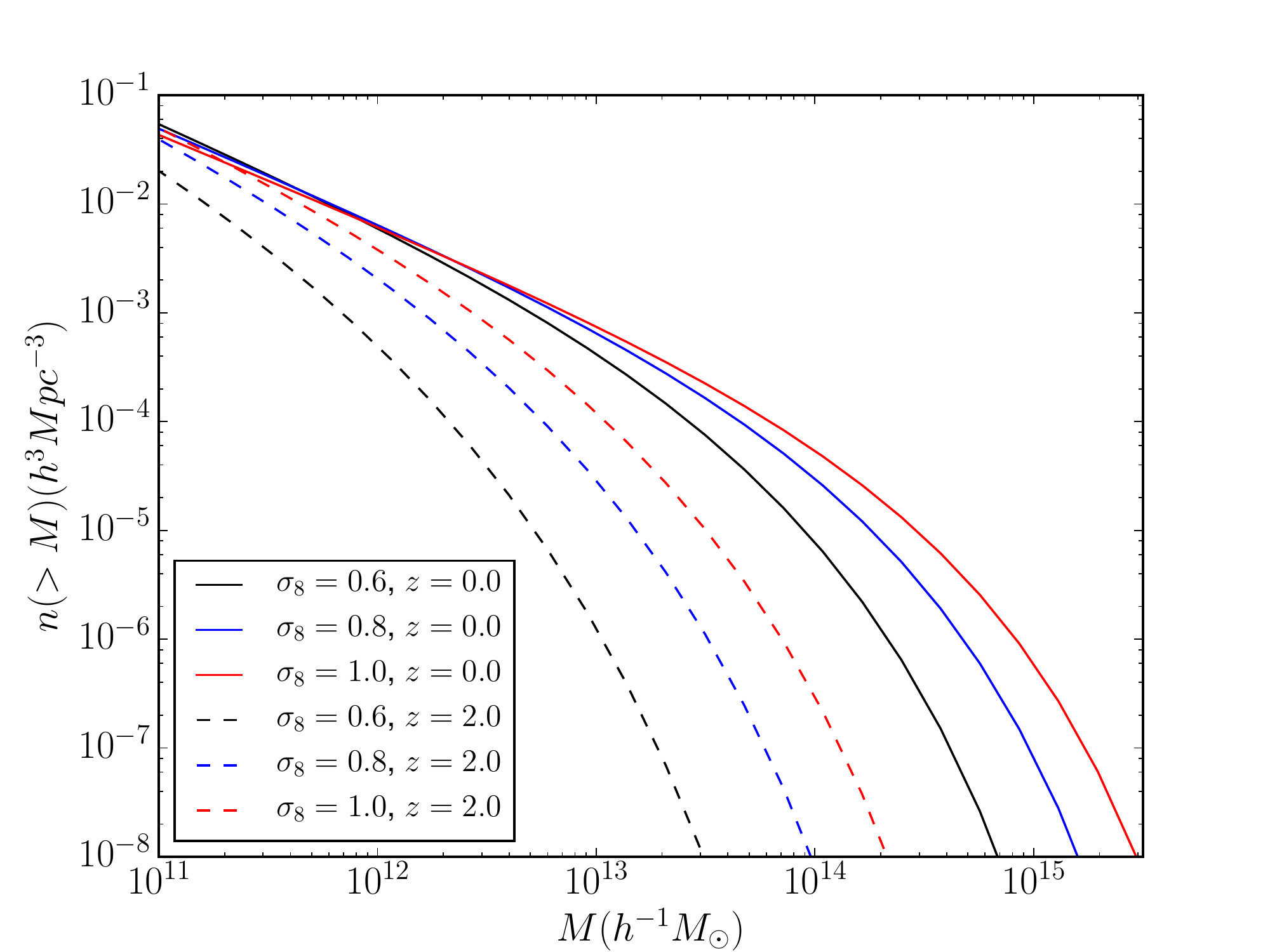}
\includegraphics[width=9cm]{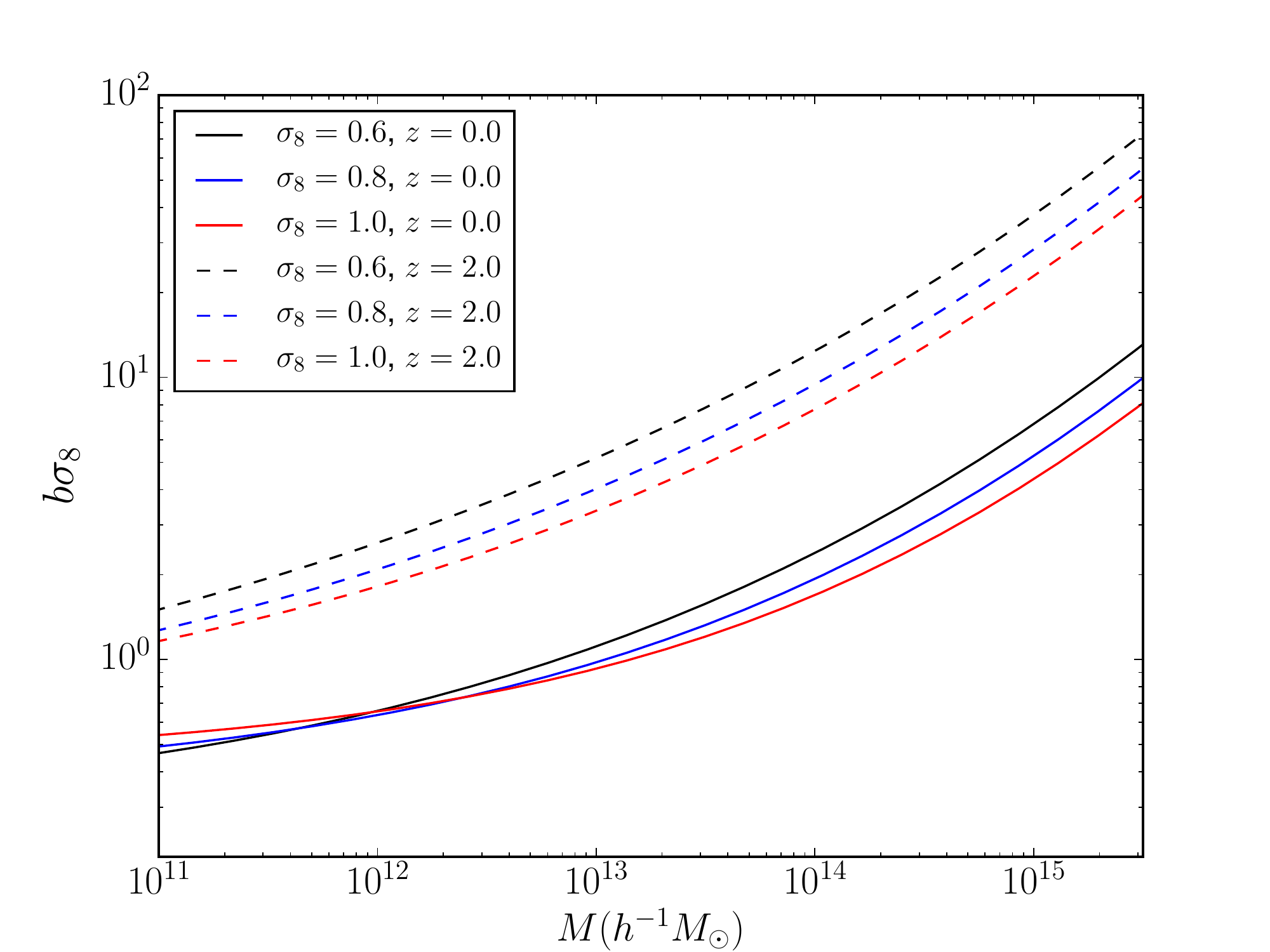}
\caption{$Top$: the number density of dark matter halo as a function of the mass threshold. $Bottom$: The clustering amplitude $b\sigma_{8}$ as a function of the halo mass. The curves with different values of $\sigma_{8}$ and redshift are also shown.}
\label{fig:nM}
\end{center}
\end{figure}

The top panel of Figure \ref{fig:nM} shows the cumulative number density of dark matter halos as a function of the mass threshold. It shows that the number density is sensitive to the change of $\sigma_{8}$ for massive halos at low redshift, and this sensitivity extends to less massive halos as the redshift increases. Therefore the massive halos at low redshift or less massive halos at higher redshift can be used to probe the asymmetric distribution of the power spectrum. Indeed, the dipole that can be measured from the abundance $N$ of the massive halos is related to the large-scale gradient in $\sigma_{8}$ (\citealt{Hirata_2009})
\begin{equation}
\mathbf{d} \propto \frac{\partial \ln{N}}{\partial\ln{\sigma_{8}} }\mathbf{p}.
\end{equation}
Under the assumption that the halo mass function just depends on $\sigma$, the studies in the context of non-Gaussianity find that (\citealt{Dalal_2008, Matarrese_2008, Slosar_2008})
\begin{equation}
\frac{\partial \ln{N}}{\partial \ln{\sigma_{8}}} = \delta_{c}(b-1),
\end{equation}
where $b$ is the linear dark matter halo bias at large scale. This model also assumes that the objects under study have a halo occupation distribution (HOD) that depends only on halo mass (\citealt{Hirata_2009}): very massive halos are highly biased and their abundance increases with $\sigma_{8}$, while low mass halos are anti-biased and their abundance decreases with $\sigma_{8}$ due to merger. In the case of occupation with only recent major mergers, this formula needs to be added with an extra suppression term $-1$ due to the merger tree dependent contribution to the large-scale bias (\citealt{Slosar_2008}).
In the following calculation, we use the linear bias arising from the excursion set approach \citep{Cole_1989, Mo_1996}:
\begin{equation}
b = 1+\frac{[\delta_{c}/\sigma(M)]^{2}-1}{\delta_{c}}.
\end{equation}
The number densities of highly biased tracers are good candidates to study the power asymmetry, because a small change in $\sigma_{8}$ can be amplified into a much larger change in its number density.

On the other hand, we note that the clustering amplitude of the galaxies is proportional to $b\sigma_{8}$. Its dependence on redshift $z$ and $\sigma_{8}$ is shown in the bottom panel of Figure \ref{fig:nM}. Regions of the Universe with different amplitudes of the power spectrum have different clustering of galaxies; these variations in clustering amplitude can be used to examine variations in the underlying power spectrum amplitude. 

As shown in Figure \ref{fig:nM}, differences in the value  of $\sigma_{8}$ can propagate into a similar fractional difference in the clustering amplitude. Because both the underlying clustering and the bias change, for most redshifts and $L_\ast$-galaxy masses increases in $\sigma_8$ lead to a decrease in clustering. Therefore in principle redshift surveys of galaxies can detect the asymmetric distribution of the power spectrum. Determining the detectability of this signal requires accounting for the uncertainties in the surveys, which is the focus of the following sections.

\section{Observability under ideal conditions}
\label{sec:ideal}

The methodology for the forecast of the uncertainty is based on the Fisher matrix formalism (\citealt{Tegmark_1997}). For the galaxy power spectrum measured in the survey, the Fisher matrix element is approximately (\citealt{Tegmark_1997b, Seo_2003})
\begin{eqnarray}
&F_{ij}=&\int_{\mathbf{k}_{\rm{min}}}^{\mathbf{k}_{\rm{max}}}\frac{\partial \ln{P(\mathbf{k})}}{\partial p_{i}}\frac{\partial \ln{P(\mathbf{k})}}{\partial p_{j}} V_{\rm{eff}}(\mathbf{k})\frac{d\mathbf{k}}{2(2\pi)^{3}}  \nonumber \\
&=&\int_{-1}^{1}\int_{k_{\rm{min}}}^{k_{\rm{max}}}\frac{\partial\ln{P(k, \mu)}}{\partial p_{i}}\frac{\partial\ln{P(k, \mu)}}{\partial p_{j}} V_{\rm{eff}}(k, \mu)\frac{2\pi k^{2}dkd\mu}{2(2\pi)^{3}}, \nonumber \\
\end{eqnarray}
where $P(\mathbf{k})$ is the observed power spectrum at mode $\mathbf{k}$, $\mu$ is the cosine of the angle between $\mathbf{k}$ and the line of sight, $p_{i}$ are the parameters to be constrained, $\mathbf{k}_{\rm{min}}$ and $\mathbf{k}_{\rm{max}}$ are defined as \cite{Zhao_2016} to only include information in the linear or quasi-linear scale (\citealt{Seo_2003})
\begin{equation}
\mathbf{k}_{\rm{min}} = \frac{2\pi}{V_{\rm{sur}}^{1/3}} [\rm{h Mpc}^{-1}], \qquad \mathbf{k}_{\rm{max}}=0.1\frac{G(0)}{G(z)} [\rm{h Mpc}^{-1}],
\end{equation}
where $G(z)$ is the growth function at redshift $z$, $V_{\rm{sur}}$ is the survey volume which is related to the effective volume of the survey by
\begin{equation}
V_{\rm{eff}}=\int\Big[\frac{n(\mathbf{r})P(k, \mu)}{n(\mathbf{r})P(k, \mu)+1}\Big]^2 d\mathbf{r}=\Big[\frac{nP(k, \mu)}{nP(k, \mu)+1}\Big]^2V_{\rm{sur}},
\end{equation}
where $n(\mathbf{r})$ is the comoving number density of galaxies at the location $\mathbf{r}$ and we assume it to be a constant in position throughout the calculation. In the limit of Gaussian likelihood surface, the expected error on the parameter $p_{i}$ is $F_{ii}^{-1/2}$ assuming the values of all other parameters are known.

In the ideal and simplest case, the galaxy power spectrum is well measured in real space, therefore we do not consider the effect from redshift space distortion, which leaves a simple relation between the power spectrum and the clustering amplitude $\partial \ln{P(k)}/\partial (\ln{b\sigma_{8}})=2$.

Given the direction of the power spectrum asymmetry measured from Planck satellite, a whole sky survey has the strongest signal when the clustering amplitudes are compared in the dipole (opposite) directions. We define a spherical coordinate system aligned with the Planck power asymmetry dipole director, with a corresponding Dipole North Pole and Dipole South Pole. 
Along the direction between the Dipole South Pole and the Dipole North Pole, the value of $\sigma_{8}$ is a linear function in comoving distance (we assume a flat Universe).
Then in a given redshift shell, $\sigma_{8}$ is a function 
only of latitude in this coordinate system. 
Therefore we can determine the value of $\sigma_{8}$ at any position in the universe with its redshift (which gives the comoving distance) and angular coordinates.

\begin{figure}[htbp]
\begin{center}
\includegraphics[width=9cm]{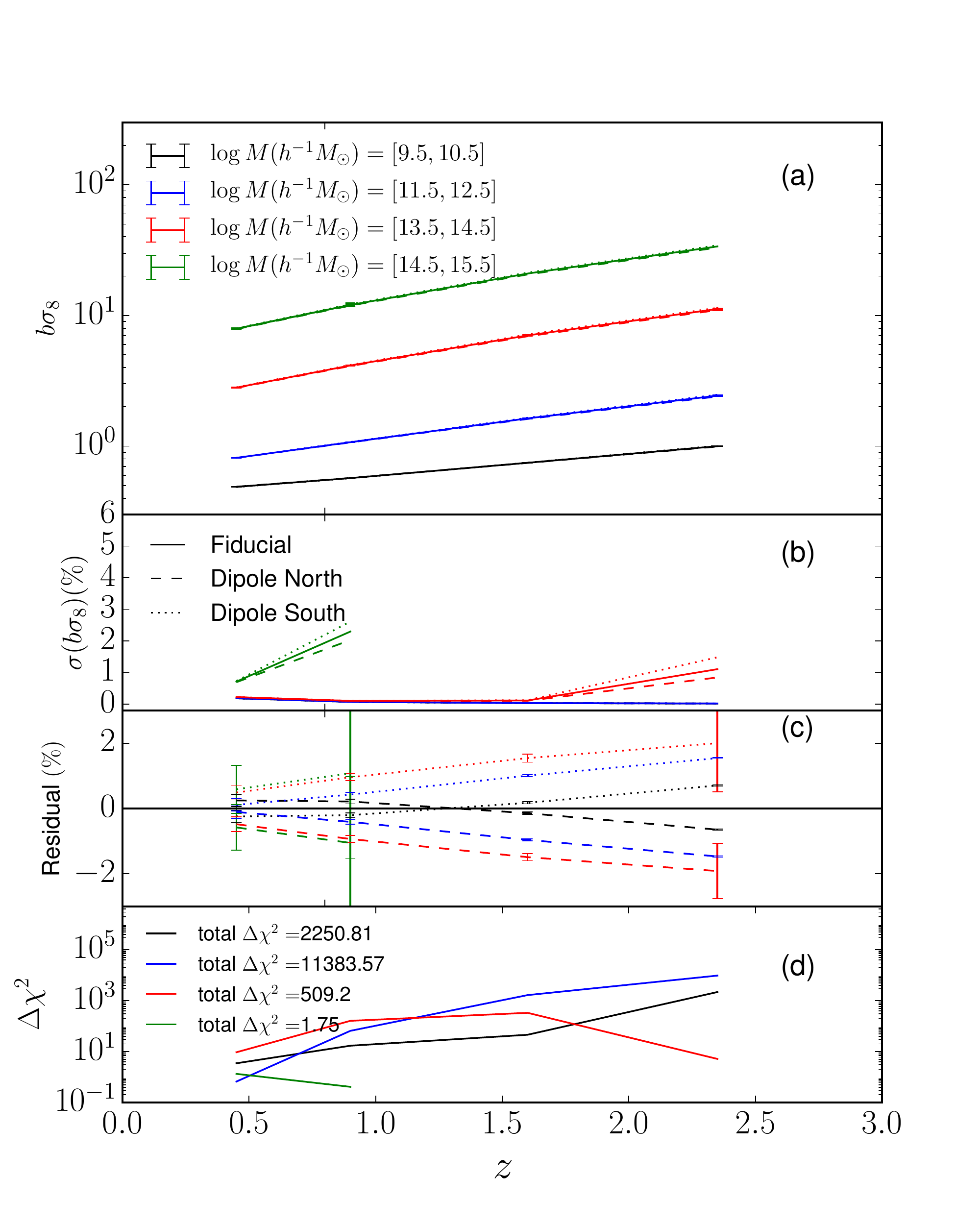}
\caption{The results for the ideal survey: (a) the clustering amplitude $b\sigma_{8}$ for halos in three different mass ranges. The results with $\sigma_{8}$ from fiducial model (solid), the mean value in the North (dashed) and the mean value in the South (dotted) are also shown. (b) The uncertainty estimated from the Fisher matrix method. (c) The fractional difference of the clustering amplitude in the North (dashed) and South (dotted) survey areas with respect to the fiducial model. The solid black line shows 0. (d) The values of $\chi^{2}$ used to measure the statistical significance for the power spectrum asymmetry in the three mass ranges. For most halo masses, results are shown in 4 redshift bins and there are correspondingly 4 degrees of freedom with respect to $\chi^2$. For halo masses in the range $14.5<\log{M (h^{-1}M_{\odot})}<15.5$, the uncertainties from Fisher matrix at high redshift is too large to be shown in the figure, therefore the error bar and $\Delta\chi^{2}$ are only shown for the first two redshift bins.}
\label{fig:ideal}
\end{center}
\end{figure}

For the ideal survey, we calculate the number density of galaxies within some mass and redshift range from the halo mass function Eq. (\ref{eq:hmf}), and assume the survey area in each hemisphere is 10,000 deg$^2$. Panel (a) in Figure \ref{fig:ideal} shows the clustering amplitude $b\sigma_{8}$ averaged over volume as a function of redshift for different halo masses. In the direction of the dipole (Dipole North Pole), the mean value of $\sigma_{8}$ for the survey is higher than the opposite direction (Dipole South Pole) and this difference increases as we go to higher redshift. The fractional difference of the clustering amplitude in these two hemispheres with respect to the fiducial model is shown in panel (c) of Figure \ref{fig:ideal}. The application of the Fisher matrix formalism gives the forecast of the uncertainties of $b\sigma_{8}$ assuming all the halos in the mass ranges can be observed, as shown in panel (b). In particular, the result is presented for four redshift bins [0.2, 0.6], [0.6, 1.2], [1.2, 2.0] and [2.0, 3.0]. Due to the large number densities and survey volumes, the uncertainties are sub-percent level for the halo masses and redshifts considered here, and these uncertainties also give the error bars in panel (c). The clustering amplitude and the uncertainty both depend on the value of $\sigma_{8}$ in the survey volume; the latter is shown in panel (b), where the effect of $\sigma_8$ on the errors are visible for the  massive halos (the red lines) at high redshift and the most massive halos (the green lines).

Considering the difference of $\sigma_{8}$ in two different hemispheres and the uncertainties constrained from the galaxy survey, we measure the statistical significance by defining a quantity
\begin{equation}
\Delta\chi^{2} = (d_{\rm NP}-d_{\rm SP})(C_{\rm NP}+C_{\rm SP})^{-1}(d_{\rm NP}-d_{\rm SP}),
\end{equation}
where $d$ is the observable (the clustering amplitude here) and $C$ is the covariance estimated from Fisher matrix. NP and SP denote the hemispheres defined by the dipole direction. $\Delta\chi^2$ is a measure of how different we expect the two hemisphere's clustering amplitude to be relative to the errors. For a given value of $\Delta\chi^2$ we can calculate the probability ($p$-value) that it would arise by chance if the clustering amplitude were homogeneous. This quantity thus tells us 
how detectable the dipole anisotropy would be if it exists.

The value of $\Delta\chi^{2}$ for the ideal survey is presented in panel (d) of Figure \ref{fig:ideal}. Due to the large number density and survey volume in ideal conditions, the uncertainty of the clustering amplitude is constrained to be very tight, which results a very significant signal for most of the halo mass ranges we considered here. 

\section{Observability for actual surveys}
\label{sec:real}

\subsection{Forecast from DESI and eROSITA}

\begin{figure}[htbp]
\begin{center}
\includegraphics[width=9cm]{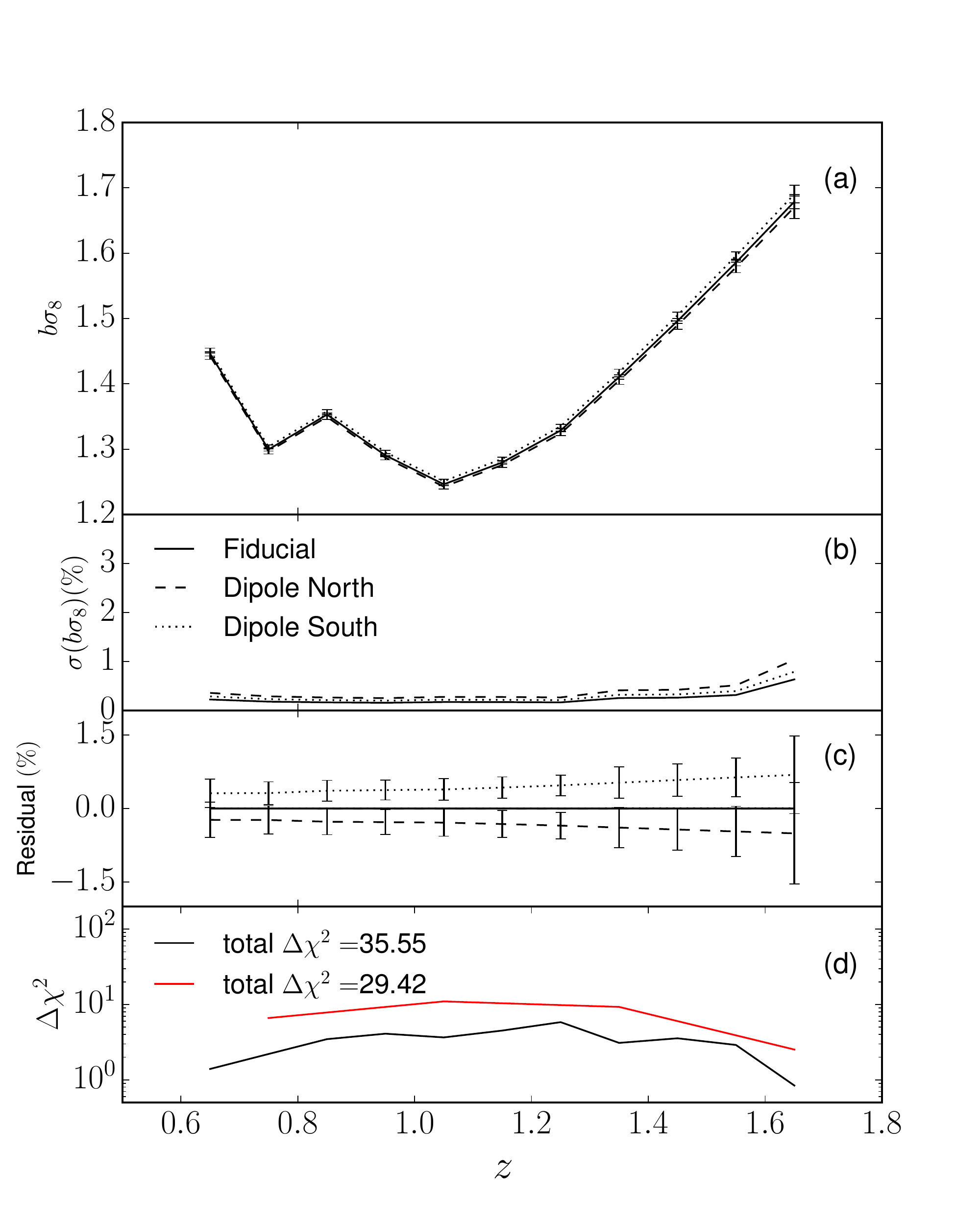}
\caption{The same as Figure \ref{fig:ideal} but for the DESI galaxy samples in 11 redshift bins and therefore 11 degrees of freedom. The red solid line is for 
different redshift binning.}
\label{fig:desi_elg_lrg}
\end{center}
\end{figure}

\begin{figure}[htbp]
\begin{center}
\includegraphics[width=9cm]{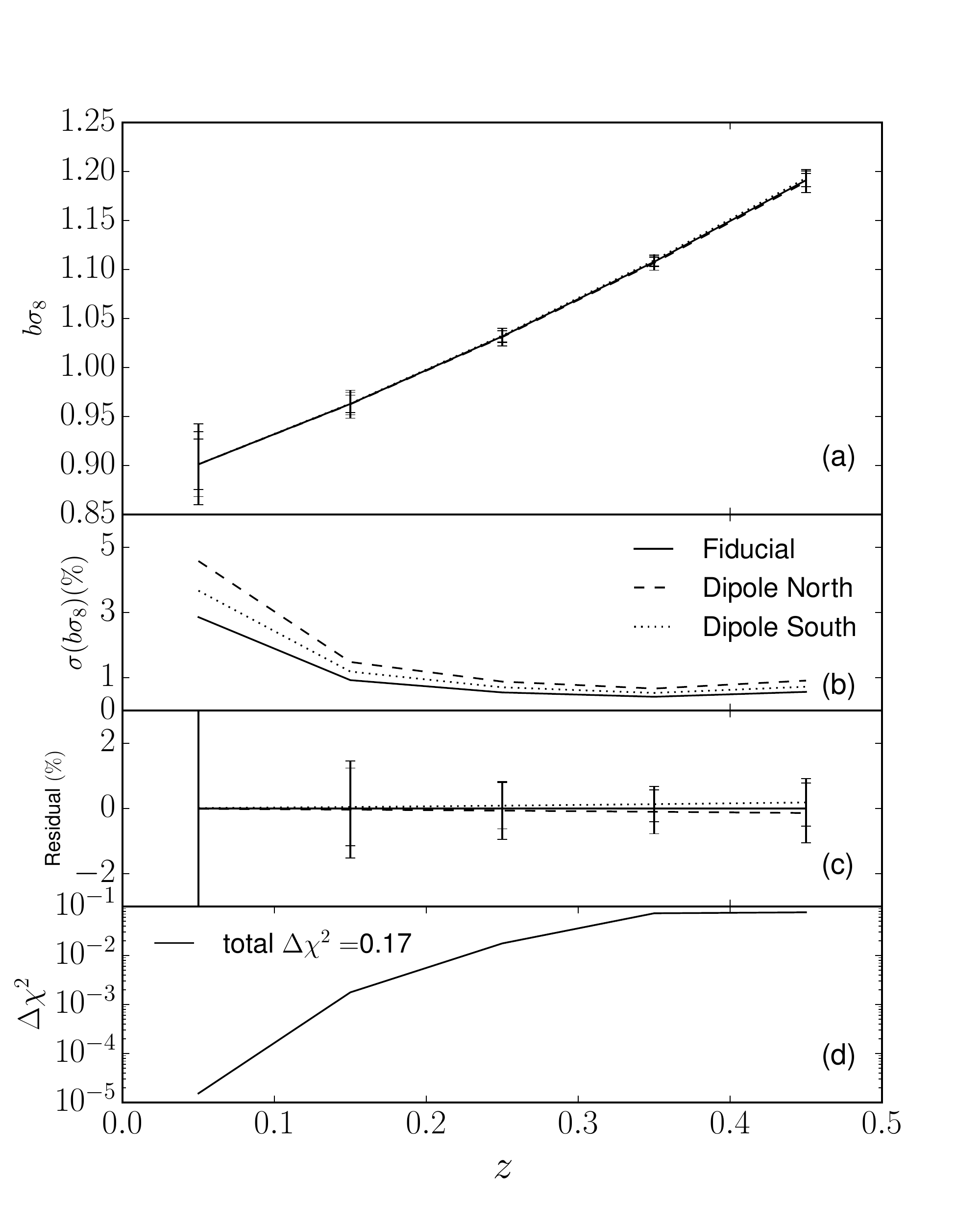}
\caption{The same as Figure \ref{fig:ideal} but for the DESI bright galaxy surveys.}
\label{fig:desi_bgs}
\end{center}
\end{figure}

\begin{figure}[htbp]
\begin{center}
\includegraphics[width=9cm]{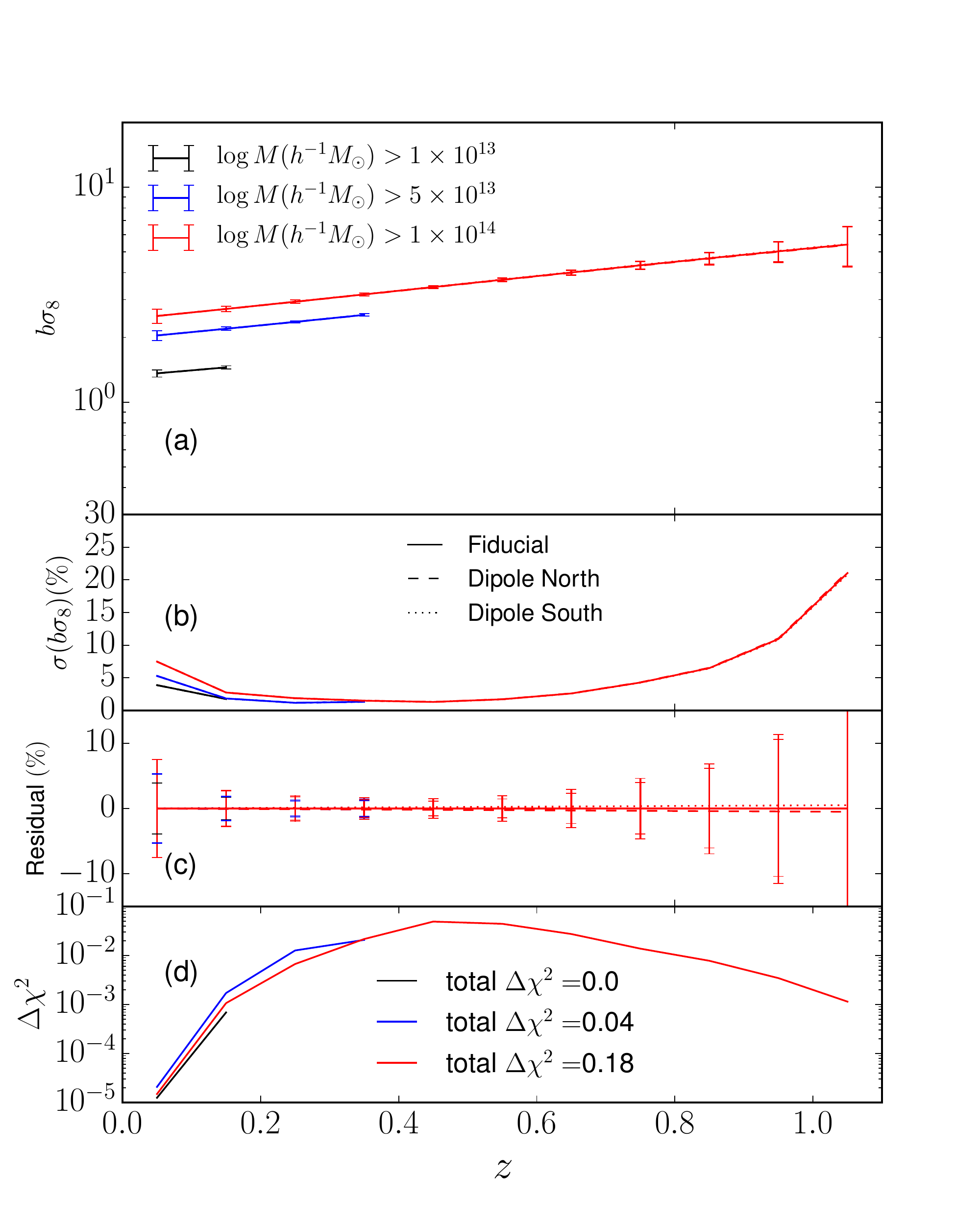}
\caption{The same as Figure \ref{fig:ideal} but for the eROSITA galaxy cluster survey. Due to the detection threshold (see Figure 3 in \citealt{eROSITA_2012}), the less massive halos are mainly in low redshift, therefore the first two mass cuts apply to galaxy clusters up to redshift 0.2 and 0.4 respectively. }
\label{fig:erosita}
\end{center}
\end{figure}

The previous section shows that the potential detectability of the power spectrum asymmetry from galaxy surveys in ideal conditions is promising. Therefore it is possible that an actual realizable survey might be able to detect this signal as well. We consider DESI (\citealt{DESI_2016}) and eROSITA (\citealt{eROSITA_2012}) and perform a similar analysis for them in this section. 

DESI is a ground based redshift survey due to begin in 2019 which has an area of 14,000 deg$^{2}$. The tracers we consider in this paper include the Bright Galaxies (BGS), Luminous Red Galaxies (LRGs) and Emission Line Galaxies (ELGs). At redshift $z>0.5$, we consider the LRG and ELG as the combined galaxy samples and for $z<0.5$, only BGS are the tracers. We use the spherical coordinate system aligned with the Planck dipole described in the previous section, 
and find that about 40\% of DESI's area is in the Dipole North Hemisphere, and about 60\% in the Dipole South Hemisphere. 

We present the analysis of the galaxy samples in Figure \ref{fig:desi_elg_lrg}. Panel (a) shows the mean value of the bias of the ELG and LRG calculated from the halo bias function. In the calculation, we assume the masses of dark matter halos that host ELGs and LRGs are $10^{12} h^{-1}M_{\odot}$ and $10^{13} h^{-1}M_{\odot}$ respectively. The other three panels have the same meaning as in the previous section. The statistical significance is again evaluated by $\Delta\chi^{2}$. Considering 11 redshift slices in the DESI galaxy survey, the value we find of $\Delta\chi^{2}/$d.o.f$\sim 35/11$ has a $p-$value $\sim 2\times10^{-4}$. We also increase the redshift binning from $\Delta z=0.1$ to $\Delta z=0.3$ and find that the $\Delta\chi^{2} /$d.o.f$\sim 29/4$ corresponding to a $p-$value $\sim 6\times10^{-6}$(the red line in Panel (d)), indicating a weak dependence of our statistic on binning choices. Both results indicate a detection above $4$--$5\sigma$. Based on Figure \ref{fig:desi_elg_lrg}, DESI can provide strong constraints on large scale gradients in the power spectrum.

The DESI BGS sample has a median redshift of $z\sim0.2$ which is much lower than ELG and LRG. Figure \ref{fig:desi_bgs} presents the results of analysis for BGS. Due to the low redshift of this sample, the difference of $\sigma_{8}$ in two hemispheres is not large. Therefore the $\chi^2$ is not as remarkable as the high redshift galaxy sample, despite the fact that the uncertainty in $\sigma_8$ is comparable.

eROSITA is an X-ray survey which is expected to detect all the galaxy clusters more massive than $3\times10^{14} h^{-1}M_{\odot}$. Following \cite{eROSITA_2012}, we assume that the fraction of eROSITA's all-sky survey usable for cluster science is $\sim0.658$, excising $\pm20$ deg around the Galactic plane. The fraction of the survey area in the two Dipole Hemispheres are the same and independent of the direction of the dipole. We present the result for eROSITA in Figure \ref{fig:erosita}. We calculate the bias of the galaxy clusters with three different mass thresholds $M_{\rm{cut}} = (1, 5, 10)\times10^{13}h^{-1}M_{\odot}$ as shown in panel (a). The uncertainty result constrained from clustering measurement shows that the survey is not able to provide robust evidence about the asymmetry of power spectrum, unless that asymmetry is much larger than the Planck estimates. The reason is two-fold: first, assuming the same mass threshold in the two survey areas, the mean mass of the halo is higher in the region with higher value of $\sigma_{8}$, which in turn has higher clustering amplitude as in Figure \ref{fig:nM}. Therefore the difference of $\sigma_{8}$ in the two hemispheres is smaller than the ones which have the same mean halo mass. Second, the number density of the galaxy clusters is not high enough, so the uncertainty estimated from the Fisher matrix formalism is quite large and dominated by significant amount of shot noise, and this effect is more important than the difference of $\sigma_{8}$ in two different regions. These factors result a low signal-to-noise measurement about the power spectrum asymmetry detection in the clustering measurement of galaxy clusters.

\subsection{Dependence on dipole amplitude}

\begin{figure}[htbp]
\begin{center}
\includegraphics[width=9cm]{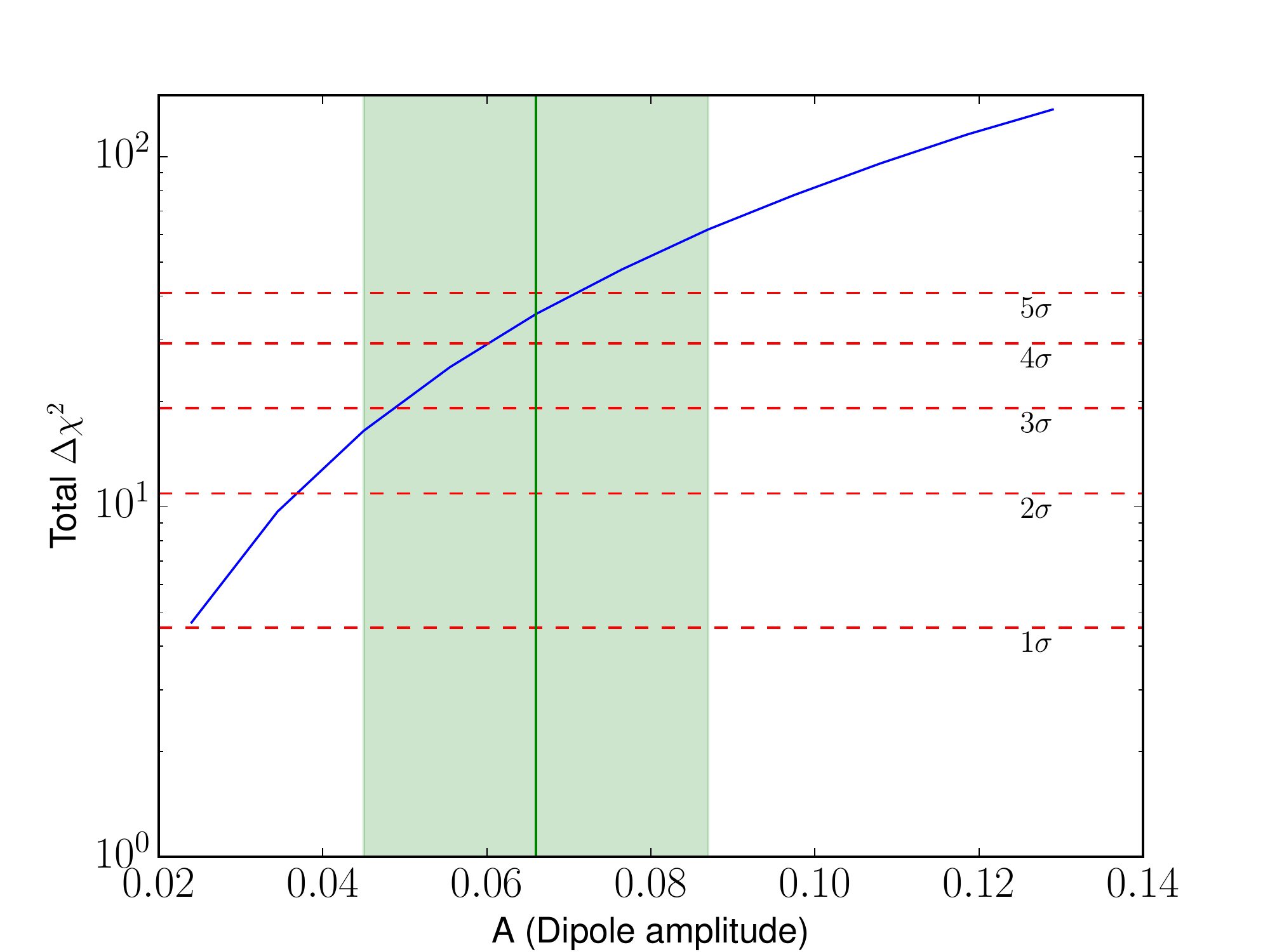}
\caption{The $\Delta\chi^{2}$ for the DESI galaxy sample as a function of the dipole amplitude. The shaded green area is the $1\sigma$ range from Planck observation, and the dashed lines correspond to the statistical significance at different levels.}
\label{fig:Amp}
\end{center}
\end{figure}

The above analysis assumes a dipole amplitude $A=0.066$; the value $A$ is what determines the gradient of $\sigma_{8}$ along the dipole direction. We investigate the dependence of the statistical significance on this value with the DESI galaxy sample. The result is presented in Figure \ref{fig:Amp}. We change the dipole amplitude from 0.024 to 0.129, while keep the model and survey information fixed. For the dipole amplitude reported from Planck, $A=0.066\pm0.021$, we find that the DESI galaxy sample can detect the signal at a level of $2.5\sigma$ with the $68\%$ lower limit of $A$, and the significance increases to higher than $5\sigma$ with the upper limit. This analysis also gives an estimate that the DESI galaxy sample can detect a $2\sigma$ signal for $A\sim0.04$.

\subsection{Dependence on scale}

\begin{figure*}[h]
\begin{center}
\includegraphics[width=8.5cm]{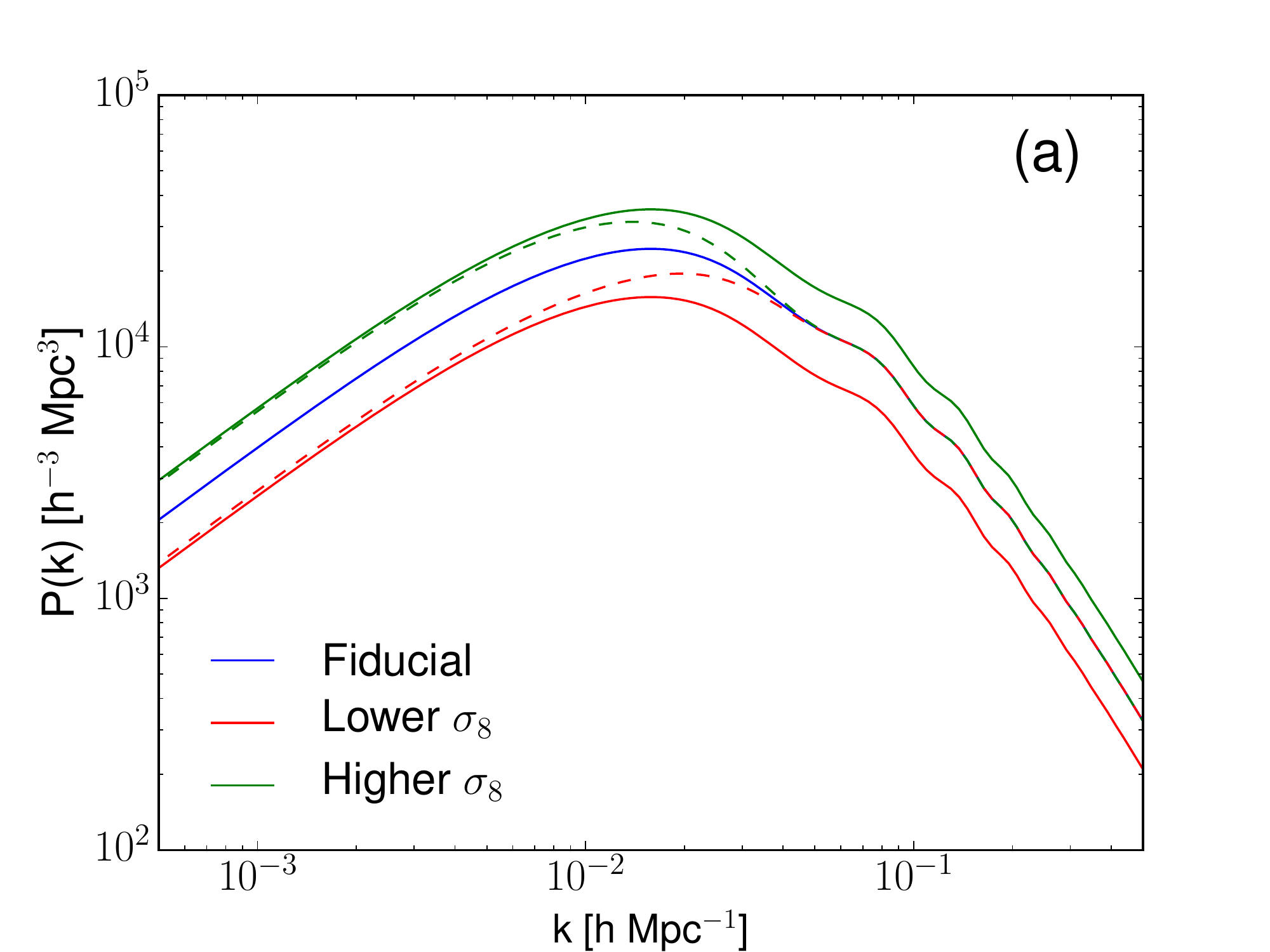}
\includegraphics[width=8.5cm]{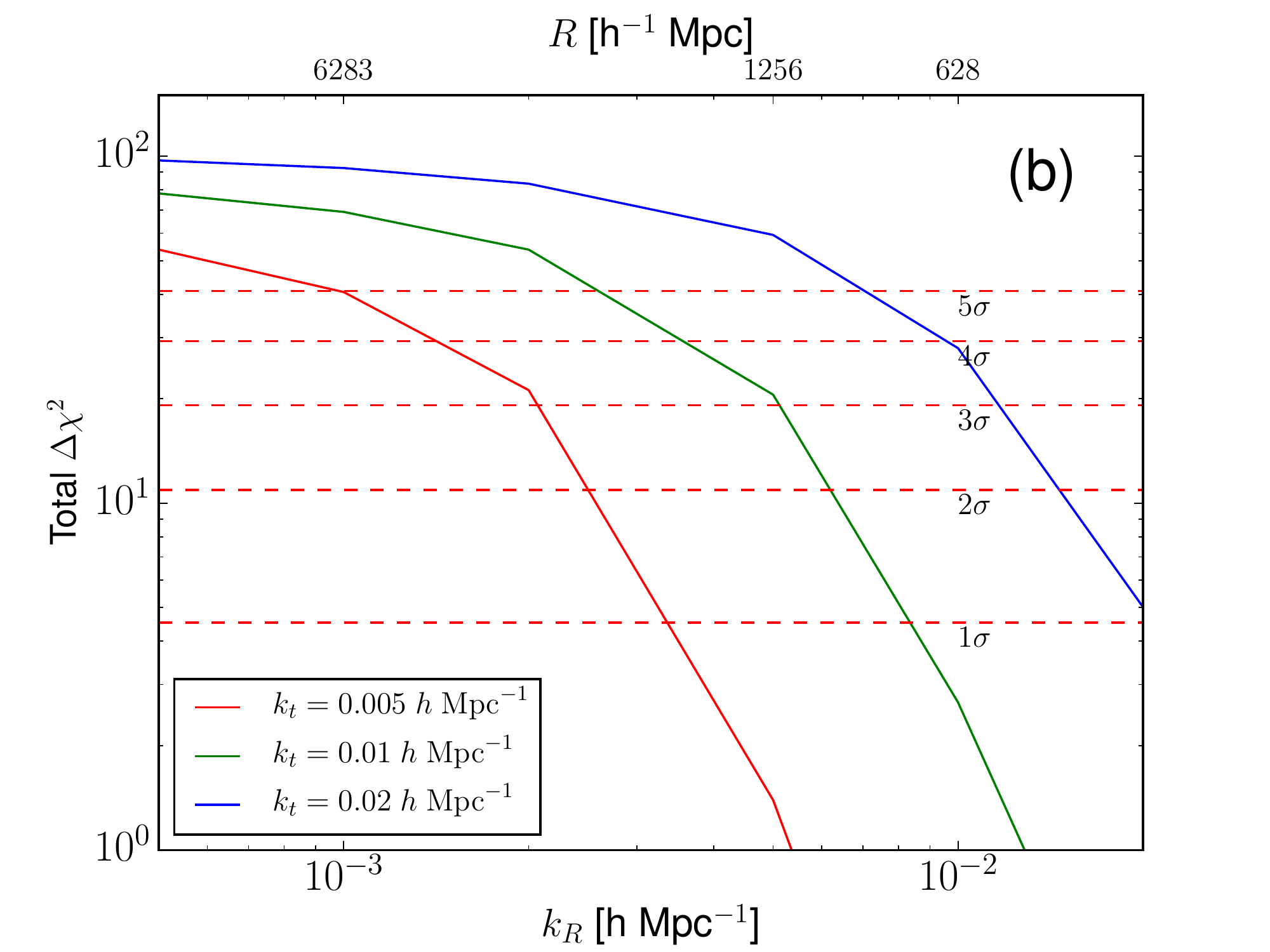}
\includegraphics[width=8.5cm]{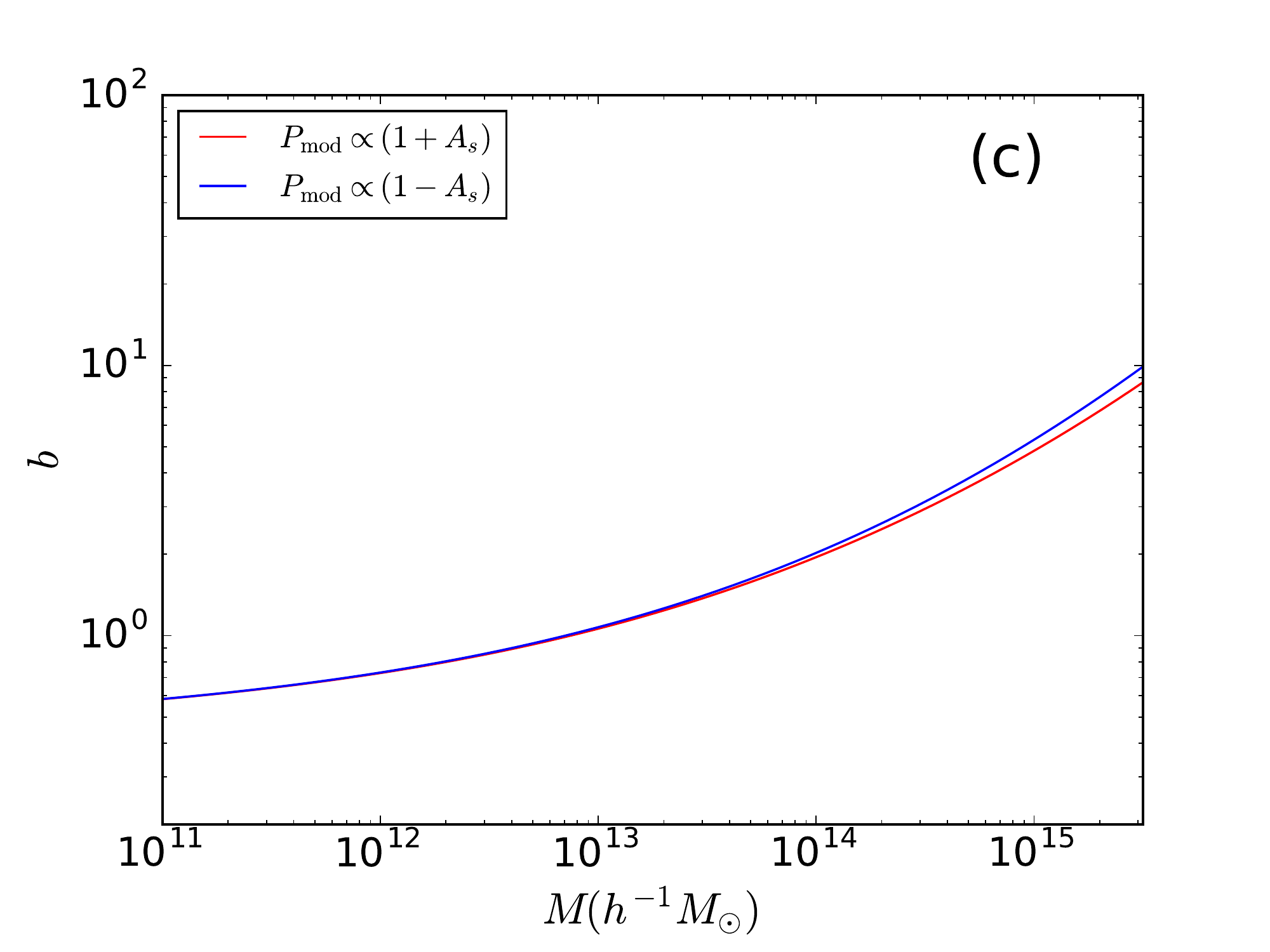}
\includegraphics[width=8.5cm]{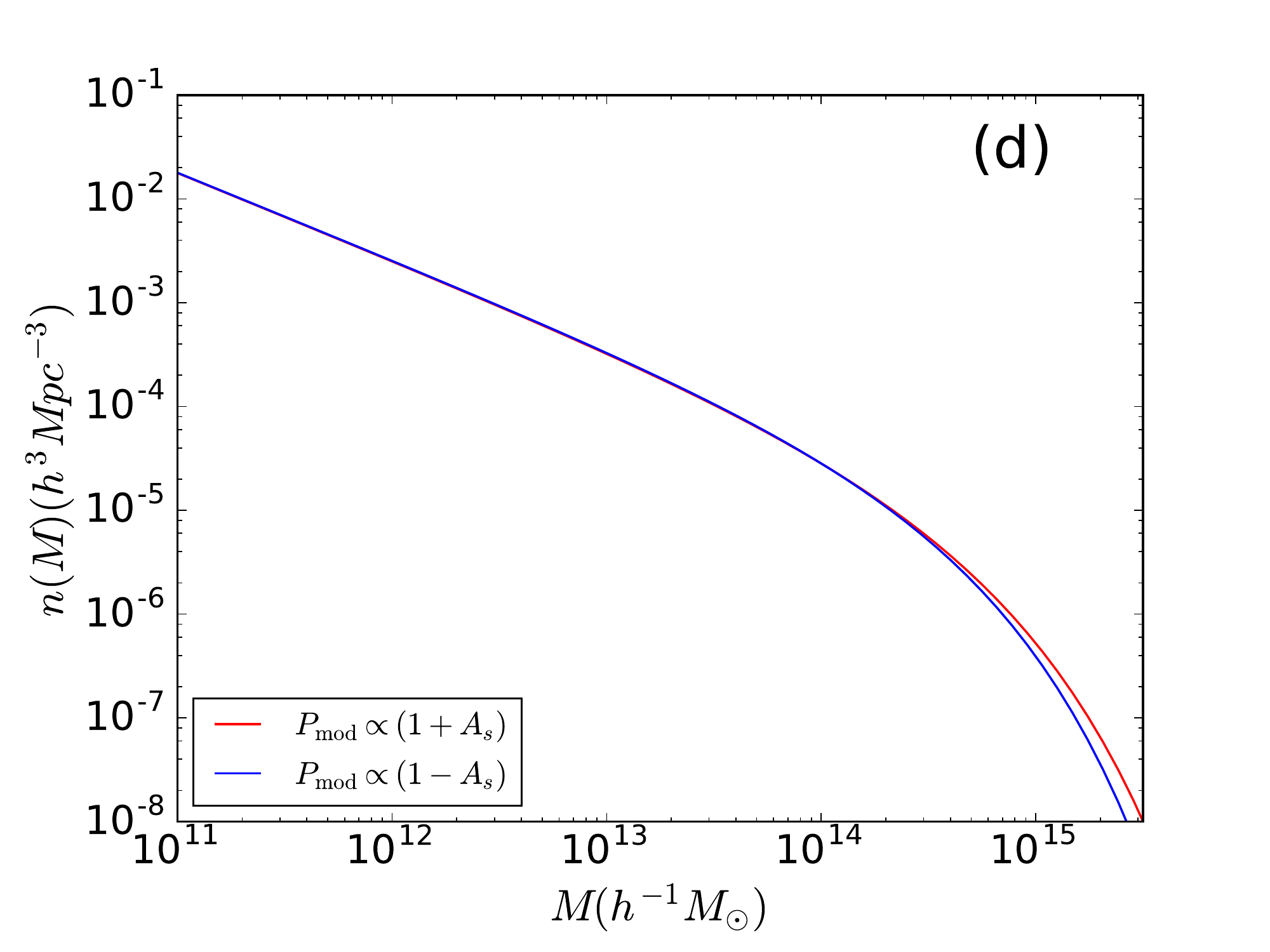}
\caption{Panel (a): The dashed lines are the modulated power spectrum of model Eq (\ref{eq:mopow}). The transition scale and width are assumed the same: $k_{t}=\sigma_{\rm{KT}}=0.02 h \rm{Mpc}^{-1}$. For the purpose of visualization, the dipole amplitude is increased by a factor of three greater than the original Planck result. The solid lines are the scale-independent power spectrum. Panel (b): The statistical significance of the power asymmetry for different models of modulation. Panel (c): the dark matter halo bias for the modulated power spectrum Eq(11), assuming a Planck-like dipole amplitude $A_{s}=0.066$, at $z=0$. The transition scale is chosen to be $k_{t}=\sigma_{\rm{KT}}=0.1 h\rm{Mpc}^{-1}$, for the mass range considered here. Panel (d): the number density of dark matter halos with the same model as Panel (c).}
\label{fig:Scale}
\end{center}
\end{figure*}

The first investigations of the hemisphere power asymmetry from CMB were performed in the low multipole range $l=2-40$ (\citealt{Hansen_2004}). Later studies found that this anomaly persists to smaller scales (\citealt{Hansen_2009,Flender_2013}). The latest research with Planck 2013 data (\citealt{Planck_2014}) claimed that the power asymmetry is significant up to $l\sim600$.  Although they did not publish a formal upper limit for the amplitude of the asymmetry, Figure 30 of \citealt{Planck_2014} suggests that it must be well below 0.01 at scales larger than that. Therefore, if the apparent hemispherical power asymmetry really exists on large scales, it must be scale dependent.

To investigate the case that the asymmetry only affects $P(k)$ on large scales, the $\sigma_{8}$ model employed in this paper needs to be modified: there is not only a spatial gradient in $\sigma_{8}$, but also a scale modulation. We thus introduce a model that leads to the modulated power spectrum (similar but more sophisticated phenomenological models can be found in \citealt{Zibin_2017, Contreras_2017}):
\begin{equation}\label{eq:mopow}
P_{\rm{mod}}(k) = P_{\rm{Fiducial}}(k)[(1\pm A_{s})^{2}(1-f(k))+f(k)],
\end{equation}
where $P_{\rm{Fiducial}}(k)$ is the power spectrum calculated from the fiducial model, $A_{s}$ is the difference between the values of $\sigma_{8}$ in the hemisphere and the fiducial model, ``+" is for the dipole North hemisphere, ``$-$" is for the dipole South hemisphere, $f(k)$ is a scale modulation and we choose it to be normalized error function
\begin{equation}
f(k) = \frac{1}{2}\left[\rm{erf}\left(\frac{k-k_{t}}{\sigma_{\rm{KT}}}\right)+1\right],
\end{equation}
where $k_{t}$ and $\sigma_{\rm{KT}}$ are the parameters related to the transition scale and width respectively. Panel (a) of Figure \ref{fig:Scale} shows an example of this modulated power spectrum, the solid lines correspond to the scale-independent power spectrum, and the dashed lines show the scale modulation. At low $k$, the power spectra have different amplitudes due to the difference of $\sigma_{8}$, this difference becomes smaller as the scale decreases and the power spectra in different dipole directions become consistent. For the purpose of visualization in the figure, the amplitude of the dipole is increased by a factor of three than the reported value from Planck. 

We apply the above modulation of the power spectrum to the DESI galaxy survey and investigate the detectability of the power asymmetry in this model. The result is presented in Panel (b) of Figure \ref{fig:Scale} for three different transition scales $k_{t}=\sigma_{\rm{KT}}=[0.005, 0.01, 0.02] h \rm{Mpc}^{-1}$, which approximately correspond to multipole moments $l\sim[50, 100, 200]$. Because the modulation of the power spectrum is scale-dependent, the observable in this frame is modified from $b\sigma_{8}$ to $b\sigma(R)$, where $R$ is the scale in configuration space which gives $k_{R}=2\pi/R$. The result shows that the clustering amplitudes of galaxies at small scales in the two opposite dipole directions are identical, since the matter power spectra at these scales have the same amplitude and the bias for the tracers is also the same. Therefore the observed clustering of galaxies at small scales is not able to probe the power asymmetry. However, for scales larger than $k_{t}$, the detectability is significant; in fact, more significant on those scales than it would have been for a scale-independent effect. The reason is that the dark matter halo bias $b$ and $\sigma(R)$ have opposite dependence on $\sigma_{8}$, which are countervailing for galaxy mass halos. When the power spectrum is modulated only on large scales, the bias $b$ in two hemispheres remains unchanged. 
This means that on large scales the galaxies trace the change in $\sigma_8$ with a constant bias, and the clustering amplitudes in two hemispheres become more different than in the scale-independent case. Therefore the significance of the potential signal is larger at these scales. 

For the modulated power spectrum Eq (11), we calculate the halo bias and number density assuming a Planck-like dipole amplitude $A_{s}=0.066$ as in Panel (c) and (d) in Figure \ref{fig:Scale}. In order to see the effect of this scale-dependent modulation in this mass range, we choose a relatively small transition scale $k_{t}=\sigma_{\rm{KT}}=0.1 h \rm{Mpc}^{-1}$. The result shows that if the spherical power asymmetry probed by CMB temperature fluctuation exists only for large scales, the dark matter halo bias and number density will remain the same everywhere.

\subsection{Constraints from the number density}

The investigation of QSO number density in \cite{Hirata_2009} places a constraint on the dipole amplitude at scales $k\sim(1.3-1.8)h \rm{Mpc}^{-1}$. Because the DESI galaxies reside in similar dark matter halos, we expect that their number densities can probe the signal at similar scales. As noted in \cite{Hirata_2009}, the null result of the dipole amplitude of  QSO number density provides a strong constraint on the inflationary explanation for the power asymmetry. 
Specifically, the calculation with QSO number density (\citealt{Hirata_2009}) finds that the dipole is no more than 0.027 at 99\% posterier probability. The measurement of the dipole amplitude in the QSO number density has an uncertainty at the level of 0.1; the conversion of this result to the dipole of the CMB power spectrum gives an uncertainty of around $\Delta A\sim0.01$. For the DESI galaxy sample considered here, we perform a similar analysis with the uncertainty of the galaxy number density (e.g. as calculated in \citealt{Yoon_2015}). For the choices appropriate for DESI for the number of sources and the shape and depth of the survey area, we find that the statistical error of the dipole of the galaxy number density is at the level of 0.003. We note that the DESI galaxy tracers do not have the high redshift and bias as the QSO tracers in \cite{Hirata_2009}, therefore the amplification of the dipole of CMB power spectrum is smaller. But it is still able to provide comparable constraint on the dipole amplitude.

On the other hand, our result shows that this dipole amplitude can be detected at better than $1\sigma$ level with galaxy clustering measurements, which can be further improved for a scale-modulated power spectrum at large scales.

\section{Discussion and Conclusion}
\label{sec:conclusions}

The anomalies in the CMB measurement reported from the WMAP and Planck satellite are intriguing and challenging to modern cosmology, if they are real features. In this paper, we focus on the dipolar asymmetry found in the CMB power spectrum, which can date back to the very early period of the universe when the primordial perturbations leave the seeds for the formation of all structure. In particular, it can be interpreted by the interaction of the inflaton field and a curvaton field. This model also predicts a large scale gradient of $\sigma_{8}$ across the universe. The implications of this model on large scale structure include a corresponding gradient in the number density of highly biased objects, which has been explored by \cite{Hirata_2009} with a null result reported, and a difference in the clustering amplitude of the large scale tracers in survey areas with different values of $\sigma_{8}$. Assuming the existence of this spatial gradient of $\sigma_{8}$ across the universe, we examined the possibility of detecting this signal from ongoing redshift surveys. 

For an ideal survey, assuming all dark matter halos above some mass can be detected, the uncertainty in the clustering measurement is quite small and thus the inferred power spectrum asymmetry can be observed with a high significance. Applying the same strategy to actual surveys like DESI and eROSITA, we find that the galaxy samples of DESI in the redshift range from 0.8 to 1.5 can yield a detection with a few $\sigma$. Note that this result is possible only when the difference of clustering amplitude due to the change of $\sigma_{8}$ is large enough relative to the uncertainties.
The BGS from DESI and galaxy clusters from eROSITA fail to satisfy these two conditions respectively, therefore they are not able to provide confirmative evidence of the power asymmetry. 

For the DESI galaxy survey, we also investigate the amplitude-dependence and scale-dependence of the statistical significance of the signal. The result reveals that this survey can probe the hemispherical power asymmetry at the $2\sigma$ level for dipole amplitude larger than $0.04$. Within the reported range of the dipole amplitude by Planck, the statistical significance can be up to $5\sigma$. 

We also modulate the model of $\sigma_{8}$ to be scale-dependent by introducing some transition scale. For scales much larger than the transition scale, the difference of clustering amplitudes in the two directions is just due to the gradient in $\sigma_{8}$, because there is no induced bias in the tracers. If these large scales can be measured at the expected statistical precision, a large scale modulation could be detected. However, for scales much smaller than the transition scale, the value of $\sigma_{8}$ and halo bias becomes identical for the two hemispheres, and there is no signal to detect. 

We note that there are limitations to our analysis. First, predicting the power spectrum of galaxy clustering requires the knowledge of the halo mass function and halo bias function, and the relation between halos and galaxies. The excursion set models used here 
are rather simplified. More accurate models calibrated with N-body simulations (\citealt{Tinker_2008, Tinker_2010}) are likely to give more accurate result. Additionally, the cosmological model we used here assumes other parameters are fixed while $\sigma_8$ varies, which may not be true for all models. We test this by changing the fiducial cosmology from a random subsample of the Planck MCMC chain (\citealt{Planck_2016_pars}), and the resulting significance of the detection for the power asymmetry is largely unchanged. We also apply a marginalized Fisher matrix formalism for the uncertainty prediction. The corresponding significance decreases and depends on the marginalized parameters and the possible priors applied.  Nevertheless, this simply tells us that the ability for DESI to detect a gradient in the power spectrum due ``specifically" to a gradient in sigma8, is less than its ability to detect a gradient in the power spectrum amplitude. Whatever the source of a gradient of the power spectrum amplitude, the forecasts based on sigma8 given in Figure \ref{fig:Amp} and \ref{fig:Scale} tell us how detectable it is. Second, we ignore corrections due to the redshift space distortions and non-linear evolution of galaxy clustering. However, considering the wide and deep survey, the uncertainties of the clustering amplitude estimated from this simplified model are constrained to be tight and a measurement of the power spectrum asymmetry is viable. Third, and perhaps most significantly, we are not accounting for systematic errors that can enter galaxy clustering measurements on large scales. The largest scales we consider here are in the regime of very weak clustering and may be susceptible to systematic effects. If these difficulties can be overcome, measurements like those described here combined with the forecast of the dipole amplitude by galaxy number density \citep{Yoon_2015}, may provide constraints on the power asymmetry and therefore inflation models.

\acknowledgements{}

This work is supported by National Science Foundation NSF-AST-1615997 and NSF-AST-1109432.

\bibliographystyle{apj}
\bibliography{anisotropy_bib}
\end{document}